\documentclass
[showpacs,amsfonts,amssymb,oneside,notitlepage,10pt,a4paper,twocolumn,preprint,balancelastpage,pra]{revtex4}%
\usepackage{amsfonts}
\usepackage{amsmath}
\usepackage{amssymb}
\usepackage{graphicx}
\usepackage[nonscript]{accents}%
\providecommand{\U}[1]{\protect\rule{.1in}{.1in}}
\begin{document}
\title{Generating highly squeezed hybrid Laguerre--Gauss modes in
large-Fresnel-number degenerate optical parametric oscillators}
\author{Carlos Navarrete-Benlloch, Germ\'{a}n J. de Valc\'{a}rcel, and Eugenio Rold\'{a}n}
\affiliation{Departament d'\`{O}ptica, Universitat de Val\`{e}ncia, Dr. Moliner 50,
46100--Burjassot, Spain}

\begin{abstract}
We theoretically describe the quantum properties of a large Fresnel number
degenerate optical parametric oscillator (DOPO) with spherical mirrors that is
pumped by a Gaussian beam and is tuned at the subharmonic frequency to a given
transverse mode family. We first analyze the classical problem and find that
only the Laguerre--Gauss modes with lowest orbital angular momentum (OAM) are
amplified above threshold. The transverse symmetry of the classically emitted
signal field depends on the family index $f$: If $f$ is even the lowest
available OAM is zero and emission occurs in a radially symmetric mode; on the
contrary, if $f$ is odd the lowest available OAM is $1$ and the emitted
signal field has the shape of a hybrid Laguerre-Gauss mode (a linear
combination of the two Laguerre-Gauss modes with OAM equal to $\pm1$) that
breaks the rotational invariance of the system. Next we focus on the squeezing
properties of this DOPO model. As for the modes with lowest OAM we demonstrate
that their quantum properties (in the linear approximation) are equal to the
standard single mode DOPO (for even $f$) or to the recently analyzed DOPO
tuned to its first transverse-mode family [Phys. Rev. Lett. \textbf{100},
203601 (2008)]. Concerning the rest of (classically empty) modes (having
larger OAM) we find that combinations of Laguerre-Gauss modes with opposite
OAM (hybrid Laguerre-Gauss modes) exhibit quadrature squeezing. This property
is independent of the even or odd character of the family index and hence has
nothing to do with the symmetry of the classically emitted signal field.
Noticeably the amount of squeezing does not depend on the pump level (it is
thus noncritical squeezing) and can be arbitrarily large for the lower OAM
nonamplified modes.

\end{abstract}

\maketitle

\section{Introduction}

Degenerate optical parametric oscillators (DOPOs) are nowadays the standard
squeezed light source. Let us remind that a light mode is said to be squeezed
if the fluctuations in one of its quadratures are below the standard quantum
limit, which is defined as the vacuum fluctuations level of that quadrature
\cite{Loudon87,WallsMilburn,Drummond04}. In DOPOs, squeezing is accomplished
thanks to the parametric down--conversion process occurring in the nonlinear
crystal together with the interference between the intracavity field and the
external vacuum fluctuations that enter into the cavity through the output
mirror \cite{Gea}. Quantum noise reductions as large as 10dB (90\%) have been
experimentally demonstrated with DOPOs \cite{Vahlbruch08}.

High precision measurements are perhaps the best known applications of
squeezed light \cite{Drummond04}, but applications to quantum information with
continuous variables are becoming increasingly important \cite{Braunstein05},
as squeezed light is the essential ingredient in generating
continuous--variable entanglement. Improving the quality and reliability of
squeezing is thus an important goal, but the generation of squeezed light with
particular spatial distributions could also be important. This has been shown
to be of utility in, e.g., high precision positioning \cite{Fabre}. Recently
Laguerre--Gauss beams are attracting much attention because of the many
potential applications of the orbital angular momentum (OAM) carried by these
light beams \cite{Arnold08}, and the generation of non classical
Laguerre--Gauss modes could thus lead to new phenomena in the interaction
between these light fields and matter.

Here we consider squeezing generation by means of DOPOs with large Fresnel
number cavities composed of spherical mirrors. In such systems, the cavity can
sustain the nonlinear interaction for several transverse modes (diffraction
losses are ideally suppressed) giving rise to new results concerning
squeezing. Quantum fluctuations in large Fresnel number cavities have been
studied in the past, and new phenomena resulting from the interplay between
quantum fluctuations and transverse pattern formation have been predicted.
Concerning cavities composed of planar mirrors, the phenomena of quantum
images \cite{Gatti} (below threshold) and the perfect noncritical squeezing of
the emerging transverse pattern linear momentum \cite{EPL,translationalPRA}
(above threshold) have been predicted. Regarding spherical mirrors, only below
threshold operation has been studied \cite{Petsas}, showing that also in this
case there appear quantum images anticipating the above threshold pattern.

Here we consider a DOPO cavity with spherical mirrors that is pumped by a
Gaussian mode. The novelty of our work with respect to previous analyses of
large Fresnel number DOPOs consists in that we study the system above its
oscillation threshold, while previous works have treated the below threshold
case \cite{Gatti,Petsas}. This has lead us to quantum fluctuations phenomena
that have not been previously described. Specifically, we shall assume that
the DOPO cavity is exactly tuned to a particular family of Laguerre--Gauss
transverse modes labeled by an integer $f$ (consisting of $f+1$
frequency-degenerate Laguerre--Gauss modes with OAMs $\pm f$, $\pm\left(
f-2\right)  $,...,$\pm l_{0}$ with $l_{0}=0$ or $1$ for even or odd $f$
respectively; this is reviewed in Appendix A). After deriving in Section II
the system's model, we first demonstrate (Section III) that the DOPO above
threshold emits a signal field consisting of pairs of photons with opposite
OAM, $+l_{0}$ and $-l_{0}$; i.e., the DOPO\ emits in the transverse mode with
the lower possible OAM, a result not derived previously. The squeezing
properties of this mode turn out to coincide with those derived in the past,
as the case $l_{0}=0$ is equivalent to the standard DOPO model
\cite{WallsMilburn}, and the case $l_{0}=1$ generalizes what has recently been
described in \cite{Rotational, Lassen09}. After reviewing the squeezing
properties of cases $f_{0}=0$ and $1$ in Subsection IV.A, we concentrate in
the squeezing properties of the rest of modes (those with $l>l_{0}$ that
remain off when the DOPO is above threshold) in Subsection IV.B. In this last
subsection we will demonstrate the main result of the present work, namely
that certain combinations of opposite OAM modes exhibit large squeezing (which
is larger for the larger values of $f$ and the lower values of $l$).
Remarkably, this squeezing is noncritical, i.e., is independent of the pumping
value and admits a simple explanation (Subsection IV.C). We find that this new
result is relevant because it establishes a simple way for generating squeezed
vacua with shapes different from the Gaussian or the TEM$_{10}$ modes. In
Section V we review the main conclusions of our work.

\section{DOPO's quantum model}

We consider a type I DOPO with spherical mirrors pumped by a coherent Gaussian
beam of frequency $2\omega_{0}$, matched to the fundamental transverse mode
$\Psi_{0}^{0}$ at that frequency. Within the cavity, pump photons are
down-converted into signal photons of frequency $\omega_{0}$ at a
$\chi^{\left(  2\right)  }$ crystal that is placed at the cavity waist. The
important assumption of this work is that the resonator is tuned so that
$\omega_{0}$ coincides with the resonance frequency of a unique transverse
mode family $f$, and is detuned far enough from any other family. Within these
conditions, the total electric field at the resonator waist plane can be
written as%
\begin{subequations}
\begin{align}
\hat{E}\left(  \mathbf{r},t\right)   &  =\hat{E}_{p}\left(  \mathbf{r}%
,t\right)  +\hat{E}_{s}\left(  \mathbf{r},t\right)  ,\\
\hat{E}_{p}\left(  \mathbf{r},t\right)   &  =i\mathcal{F}_{p}\hat{A}%
_{p}\left(  \mathbf{r},t\right)  e^{-2i\omega_{0}t}+\mathrm{H.c.},\\
\hat{E}_{s}\left(  \mathbf{r},t\right)   &  =i\mathcal{F}_{s}\hat{A}%
_{s}\left(  \mathbf{r},t\right)  e^{-i\omega_{0}t}+\mathrm{H.c.},
\end{align}
where we have introduced the single-photon field voltages%
\end{subequations}
\begin{equation}
\mathcal{F}_{p}=\sqrt{2}\mathcal{F}_{s}=\sqrt{\frac{2\hbar\omega_{0}%
}{\varepsilon_{0}n_{c}L_{\mathrm{eff}}}},
\end{equation}
being $n_{c}$ is the crystal refractive index and $L_{\mathrm{eff}}$ the
effective resonator length (see Appendix A), and the slowly varying envelopes
\begin{subequations}
\label{SVEs}%
\begin{align}
\hat{A}_{p}\left(  \mathbf{r},t\right)   &  =\hat{a}_{00}\left(  t\right)
\Psi_{0}^{0}\left(  \mathbf{r}\right)  ,\label{Ap}\\
\hat{A}_{s}\left(  \mathbf{r},t\right)   &  =%
%TCIMACRO{\dsum _{l}}%
%BeginExpansion
{\displaystyle\sum_{l}}
%EndExpansion
\frac{1}{\left(  1+\delta_{0l}\right)  }\left[  \hat{a}_{+l}\left(  t\right)
\Psi_{\left(  f-l\right)  /2}^{+l}\left(  \mathbf{r}\right)  \right.
\label{As}\\
&  \left.  +\hat{a}_{-l}\left(  t\right)  \Psi_{\left(  f-l\right)  /2}%
^{-l}\left(  \mathbf{r}\right)  \right]  ,\nonumber
\end{align}
where the interaction picture boson operators obey the canonical commutation
relations $\left[  \hat{a}_{m}\left(  t\right)  ,\hat{a}_{n}^{\dagger}\left(
t\right)  \right]  =\delta_{mn}$, the Kronecker symbol $\delta_{0l}$ is
introduced because $\hat{a}_{-0}\left(  t\right)  =\hat{a}_{+0}\left(
t\right)  $, and $\Psi_{\left(  f-l\right)  /2}^{\pm l}\left(  \mathbf{r}%
\right)  $ are the resonator Laguerre--Gauss modes at the resonator waist
plane. The expressions for these modes as well as some of their properties are
given in Appendix A. Here we just note that%
\end{subequations}
\begin{equation}
\Psi_{n}^{\pm l}\left(  \mathbf{r}\right)  =\mathcal{N}_{n}^{l}u_{n}%
^{l}\left(  r\right)  \exp\left(  \pm il\phi\right)  , \label{modetes}%
\end{equation}
where $\mathbf{r}=\left(  r\cos\phi,r\sin\phi\right)  $ are the resonator
waist plane cartesian coordinates ($r$ and $\phi$ are the corresponding polar
coordinates), $\mathcal{N}_{n}^{l}$ is a normalization factor,%
\begin{equation}
u_{n}^{l}\left(  r\right)  =\frac{1}{w}\left(  \frac{\sqrt{2}r}{w}\right)
^{l}L_{n}^{l}\left(  \frac{2r^{2}}{w^{2}}\right)  e^{-r^{2}/w^{2}},
\end{equation}
$L_{n}^{l}$ is the modified Laguerre polynomial, and $w$ is the beam waist
radius (that depends on frequency). In (\ref{As}) the sum in $l$ runs along
all the members of the transverse mode family $f$ [$l\in\left\{
f,f-2,...,l_{0}\right\}  $] and this will hold for all sums in $l$ along the
rest of the article. Finally, note that the Laguerre-Gauss modes appearing in
(\ref{Ap}) and (\ref{As}) are evaluated at pump and signal frequencies,
respectively, which determine the beam waist radius $w$ (see Appendix A).

For the sake of later use we mention here that instead of the Laguerre-Gauss
modes one can use the hybrid modes, defined (for $l\neq0$) as
\begin{align}
H_{\mathrm{c},n}^{l}\left(  \mathbf{r}\right)   &  =\frac{1}{\sqrt{2}}\left[
\Psi_{n}^{+l}\left(  \mathbf{r}\right)  +\Psi_{n}^{-l}\left(  \mathbf{r}%
\right)  \right]  =\sqrt{2}\mathcal{N}_{n}^{l}u_{n}^{l}\left(  r\right)
\cos\left(  l\phi\right)  ,\\
H_{\mathrm{s},n}^{l}\left(  \mathbf{r}\right)   &  =\frac{1}{\sqrt{2}i}\left[
\Psi_{n}^{+l}\left(  \mathbf{r}\right)  -\Psi_{n}^{-l}\left(  \mathbf{r}%
\right)  \right]  =\sqrt{2}\mathcal{N}_{n}^{l}u_{n}^{l}\left(  r\right)
\sin\left(  l\phi\right)  ,
\end{align}
(see Appendix A) in order to expand the slowly varying envelope of the signal
field, whose corresponding boson operators relate to the Laguerre-Gauss mode
ones as
\begin{subequations}
\label{GLtoH}%
\begin{align}
\hat{a}_{\mathrm{c},l}  &  =\left[  \hat{a}_{+l}\left(  t\right)  +\hat
{a}_{-l}\left(  t\right)  \right]  /\sqrt{2},\\
\hat{a}_{\mathrm{s},l}  &  =i\left[  \hat{a}_{+l}\left(  t\right)  -\hat
{a}_{-l}\left(  t\right)  \right]  /\sqrt{2}.
\end{align}
For $l=0$ we still use the Laguerre--Gauss mode $\Psi_{n}^{0}\left(
\mathbf{r}\right)  $.

In the interaction picture, assuming perfect phase--matching as well as exact
resonance between the fields' frequencies and the cavity resonances, the
Hamiltonian of the system is $\hat{H}=\hat{H}_{\mathrm{ext}}+\hat
{H}_{\mathrm{int}}$, with
\end{subequations}
\begin{subequations}
\label{Hamiltonian}%
\begin{align}
\hat{H}_{\mathrm{ext}}  &  =i\hbar\mathcal{E}_{p}\left(  \hat{a}_{00}%
^{\dagger}-\hat{a}_{00}\right)  ,\\
\hat{H}_{\mathrm{int}}  &  =i\hbar%
%TCIMACRO{\dsum \limits_{l}}%
%BeginExpansion
{\displaystyle\sum\limits_{l}}
%EndExpansion
\frac{\chi_{l}}{1+\delta_{0l}}\left(  \hat{a}_{00}\hat{a}_{+l}^{\dagger}%
\hat{a}_{-l}^{\dagger}-\hat{a}_{00}^{\dagger}\hat{a}_{+l}\hat{a}_{-l}\right)
,
\end{align}
where $\hat{H}_{\mathrm{ext}}$ describes the external pumping process and
$\hat{H}_{\mathrm{int}}$ describes the down-conversion process occurring in
the $\chi^{\left(  2\right)  }$ crystal. Notice that OAM conservation imposes
the creation/annihilation of a pair of signal photons each with an opposite
value of $l$.

The nonlinear coupling constants $\chi_{l}$ in $\hat{H}_{\mathrm{int}}$ read
\end{subequations}
\begin{equation}
\chi_{l}=12\frac{\chi^{\left(  2\right)  }l_{c}}{w_{p}}\left(  \frac
{\omega_{0}}{n_{c}L_{\mathrm{eff}}}\right)  ^{3/2}\sqrt{\frac{\hbar}%
{\pi\varepsilon_{0}}}I_{l}, \label{coupling}%
\end{equation}
with $\chi^{\left(  2\right)  }$ the second order susceptibility of the
nonlinear crystal whose thickness is $l_{c}$, $w_{p}$ the beam spot size at
the pump frequency, and%
\begin{equation}
I_{l}=\frac{\left[  \left(  f-l\right)  /2\right]  !}{\left[  \left(
f+l\right)  /2\right]  !}\int_{0}^{+\infty}due^{-2u}u^{l}\left[  L_{\left(
f-l\right)  /2}^{l}\left(  u\right)  \right]  ^{2}, \label{IL}%
\end{equation}
which are proportional to the overlapping integrals between the three modes
involved in the particular parametric process. This means that the nonlinear
coupling between pump and signal photons is larger the lower is the OAM of the
latter ones, i.e., $\chi_{f}<\chi_{f-2}<...<\chi_{l_{0}}$. This property will
play an important role as we show below. Let us finally remark the the pump
parameter $\mathcal{E}_{p}$ is proportional to the external pump amplitude and
is taken as real without loss of generality.

We will be interested in calculating normally ordered correlations of
different mode operators, and to do so we use the generalized $P$
representation and its equivalent set of Langevin equations \cite{Drummond80}.
In this representation to every pair of boson operators $\left(  \hat{a}%
_{j},\hat{a}_{j}^{\dagger}\right)  $ it corresponds a pair of independent
stochastic amplitudes $\left(  \alpha_{j},\alpha_{j}^{+}\right)  $ that are
complex-conjugated in average, i.e., $\left\langle \alpha_{j}^{+}\right\rangle
=\left\langle \alpha_{j}\right\rangle ^{\ast}$. The stochastic average of any
function of $\left(  \alpha_{j},\alpha_{j}^{+}\right)  $ equals the
corresponding normally ordered, quantum mechanical expected value. The
equations of evolution of these amplitudes are derived in Appendix B by
following the standard procedure \cite{Carmichael, Gardiner00}. Assuming that
losses occur just at one of the cavity mirrors at rates $\gamma_{p}$ for the
pump mode and $\gamma_{s}$ for all possible transverse signal modes (hence the
assumption of a large Fresnel number resonator) the Langevin equations of the
system read
\begin{subequations}
\label{Langevin}%
\begin{align}
\dot{\alpha}_{00}  &  =\mathcal{E}_{p}-\gamma_{p}\alpha_{00}-%
%TCIMACRO{\dsum \limits_{l}}%
%BeginExpansion
{\displaystyle\sum\limits_{l}}
%EndExpansion
\frac{\chi_{l}}{1+\delta_{0l}}\alpha_{+l}\alpha_{-l},\\
\dot{\alpha}_{00}^{+}  &  =\mathcal{E}_{p}\mathcal{-}\gamma_{p}\alpha_{00}%
^{+}\mathcal{-}%
%TCIMACRO{\dsum \limits_{l}}%
%BeginExpansion
{\displaystyle\sum\limits_{l}}
%EndExpansion
\frac{\chi_{l}}{1+\delta_{0l}}\alpha_{+l}^{+}\alpha_{-l}^{+},\\
\dot{\alpha}_{\pm l}  &  =-\gamma_{s}\alpha_{\pm l}+\chi_{l}\alpha_{\mp l}%
^{+}\alpha_{00}+\sqrt{\chi_{l}\alpha_{00}}\xi_{\pm l}\left(  t\right)  ,\\
\dot{\alpha}_{\pm l}^{+}  &  =-\gamma_{s}\alpha_{\pm l}^{+}+\chi_{l}%
\alpha_{\mp l}\alpha_{00}^{+}+\sqrt{\chi_{l}\alpha_{00}^{+}}\xi_{\pm l}%
^{+}\left(  t\right)  ,
\end{align}
where $l\in\left\{  f,f-2,...,l_{0}\right\}  $. The noises $\left(  \xi
_{l},\xi_{l}^{+}\right)  $ are independent complex Gaussian white noise
sources verifying $\left\langle \xi_{l}\left(  t\right)  \right\rangle
=\left\langle \xi_{l}^{+}\left(  t\right)  \right\rangle =0$ and%
\end{subequations}
\begin{equation}
\left\langle \xi_{l}\left(  t\right)  \xi_{l^{\prime}}^{\ast}\left(
t^{\prime}\right)  \right\rangle =\left\langle \xi_{l}^{+}\left(  t\right)
\left[  \xi_{l^{\prime}}^{+}\left(  t^{\prime}\right)  \right]  ^{\ast
}\right\rangle =\delta_{ll^{\prime}}\delta\left(  t-t^{\prime}\right)  ,
\label{c-corr}%
\end{equation}
being null the rest of correlations. The rest of noises verify $\xi
_{-l}\left(  t\right)  =\xi_{l}^{\ast}\left(  t\right)  $ and $\xi_{-l}%
^{+}\left(  t\right)  =\left[  \xi_{l}^{+}\left(  t\right)  \right]  ^{\ast}$

Within the generalized $P$-representation, we define the amplitude and phase
quadratures, respectively, of a mode $j$ as $X_{j}=\left(  \alpha_{j}%
^{+}+\alpha_{j}\right)  $ and $Y_{j}=i\left(  \alpha_{j}^{+}-\alpha
_{j}\right)  $. Outside the cavity, it is measured the variance spectrum of
quadrature $X_{j}$ (analogously for $Y_{j}$), which can be calculated as
$V_{x,j}^{\mathrm{out}}\left(  \omega\right)  =1+S_{x,j}^{\mathrm{out}}\left(
\omega\right)  $, with the squeezing spectrum given, as a function of the
intracavity amplitudes, by \cite{Gea}%
\begin{equation}
S_{x,j}^{\mathrm{out}}\left(  \omega\right)  =2\gamma_{j}\int_{-\infty
}^{+\infty}d\tau e^{-i\omega\tau}\left\langle X_{j}\left(  t\right)
,X_{j}\left(  t+\tau\right)  \right\rangle , \label{Sout}%
\end{equation}
where the factor $2\gamma_{j}$ comes from the input-output relations
\cite{Collet}. We use the notation $\left\langle a,b\right\rangle
=\left\langle ab\right\rangle -\left\langle a\right\rangle \left\langle
b\right\rangle $. Notice that $S_{x,j}^{\mathrm{out}}\left(  \omega
_{s}\right)  =-1$ signals perfect squeezing outside the cavity for quadrature
$X_{j}$ at detection frequency $\omega_{s}$ (which must not be confused with
the optical frequency, as it has contributions of every pair of modes lying in
opposite sidebands around the optical frequency $\omega_{j}+\omega_{s}$, where
$\omega_{j}$ is the carrier frequency of the detected mode \cite{Gea}). Hence,
by solving the Langevin equations for the amplitudes $\alpha_{j}$ in terms of
the noises $\xi_{k}$, one is able to obtain the variances of the quadratures
involved in the problem, and thus the squeezing properties of the field.

\section{Classical emission}

The classical equations describing the field inside the DOPO are obtained from
the quantum Langevin ones by identifying the stochastic amplitudes $\alpha
_{j}$ with the classical normal variables of each mode, by making $\alpha
_{j}^{+}\rightarrow\alpha_{j}^{\ast}$, and by neglecting noise terms. They
read
\begin{subequations}
\label{classical eqs}%
\begin{align}
\dot{\alpha}_{00}  &  =\mathcal{E}_{p}-\gamma_{p}\alpha_{00}-%
%TCIMACRO{\dsum \limits_{l}}%
%BeginExpansion
{\displaystyle\sum\limits_{l}}
%EndExpansion
\frac{\chi_{l}}{1+\delta_{0l}}\alpha_{+l}\alpha_{-l}%
,\label{classical eqs pump}\\
\dot{\alpha}_{\pm l}  &  =-\gamma_{s}\alpha_{\pm l}+\chi_{l}\alpha_{\mp
l}^{\ast}\alpha_{00}. \label{classical eqs l}%
\end{align}

Equations (\ref{classical eqs}) have two types of stationary solutions. If we
define the normalized pump parameter as%
\end{subequations}
\begin{equation}
\sigma=\frac{\chi_{l_{0}}}{\gamma_{p}\gamma_{s}}\mathcal{E}_{p}, \label{sigma}%
\end{equation}
it is easy to prove that the below-threshold solution%
\begin{equation}
\bar{\alpha}_{00}=\mathcal{E}_{p}/\gamma_{p},\ \ \ \ \ \bar{\alpha}%
_{l}=0\ \ \forall l, \label{stationary below}%
\end{equation}
is stable for $\sigma<1$ and unstable for $\sigma>1$ ($\sigma=1$ thus defines
the classical threshold for emission). Apart from this \textit{trivial}
solution, there are $\left(  f-l_{0}\right)  /2+1$ possible stationary
solutions in which the signal field is nonzero. The form of these solutions
is
\begin{subequations}
\label{stationary above}%
\begin{align}
\bar{\alpha}_{00}  &  =\gamma_{s}/\chi_{k},\text{ \ \ \ }\bar{\alpha}_{\pm
l}=0\text{ \ }\forall\text{ }l\neq k,\\
\bar{\alpha}_{\pm k}  &  =\rho_{k}e^{\mp i\theta_{k}},\ \ \ \rho_{k}^{2}%
\equiv\frac{1+\delta_{0,k}}{g^{2}\kappa_{k}}\left(  \sigma-\frac{1}{\kappa
_{k}}\right)  ,
\end{align}
with%
\end{subequations}
\begin{equation}
g=\frac{\chi_{l_{0}}}{\sqrt{\gamma_{p}\gamma_{s}}},\ \ \ \kappa_{k}=\frac
{\chi_{l}\bar{\alpha}_{00}}{\gamma_{s}}=\frac{\chi_{k}}{\chi_{l_{0}}}%
=\frac{I_{k}}{I_{l_{0}}}, \label{parameters}%
\end{equation}
see Eqs. (\ref{coupling}) and (\ref{IL}). Note that in any of these solutions
only two opposite OAM values are excited ($\pm k$), the rest remaining below
threshold. As for the phases $\theta_{k}$, $\theta_{0}=0$ and $\theta_{k\neq
0}=\theta$ is arbitrary. This arbitrariness appears because Eqs.
(\ref{classical eqs}) have the symmetry $\alpha_{\pm l}\rightarrow\alpha_{\pm
l}\exp\left(  \pm i\beta\right)  $, reflecting the rotational invariance of
the system, which leaves undefined the phase difference between opposite OAM modes.

\begin{figure}[t]

\includegraphics[
width=3.3in
]%
{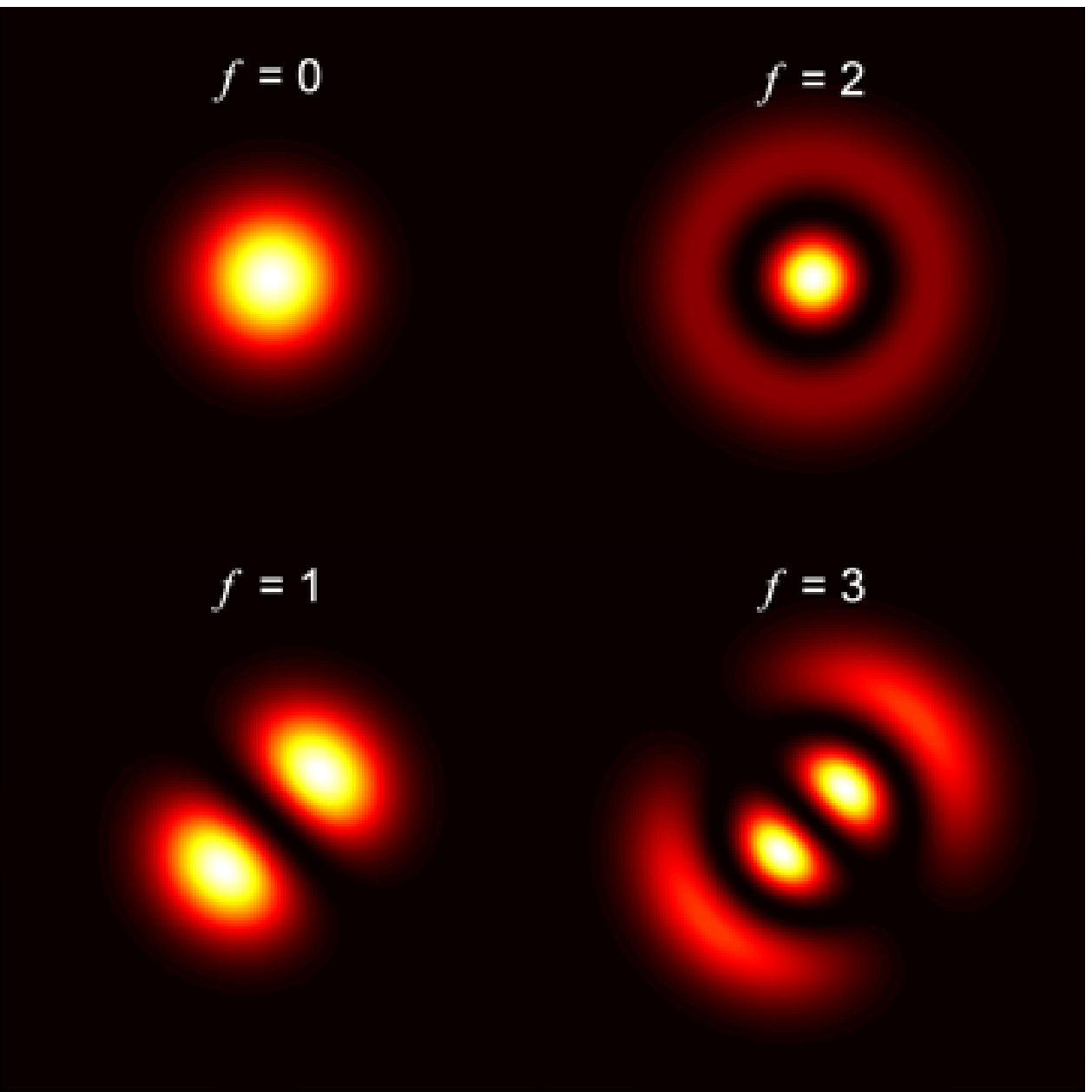}%
\\
{\small Figure 1.- Transverse profile of the signal field modulus above
threshold.}%

\end{figure}

As for the stability, it is easy to show that the only solution that is
linearly stable above threshold is precisely that in which the lower OAM modes
are switched on, i.e., Eq. (\ref{stationary above}) for $k=l_{0}$. These are,
in fact, the solutions with the lowest threshold $\sigma=1$ [see Eq.
(\ref{stationary above})], as $\chi_{l_{0}}>\chi_{l}$ as already stressed [see
Eq. (\ref{coupling})]. Plugging Eq. (\ref{stationary above}) into Eq.
(\ref{As}) one finds the slowly varying amplitudes of these stable classical
solutions as%
\begin{subequations}
\begin{align}
\bar{A}_{s}\left(  \mathbf{r}\right)   &  =\rho_{0}\Psi_{f/2}^{0}\left(
\mathbf{r}\right)  \text{ for even }f\\
\bar{A}_{s}\left(  \mathbf{r}\right)   &  =\sqrt{2}\rho_{1}H_{\mathrm{c}%
,\left(  f-1\right)  /2}^{1}\left(  r,\phi-\theta\right)  \text{ for odd }f
\label{As1}%
\end{align}
with $\rho_{j}$ given by Eq. (\ref{stationary above}). In Fig. 1 the square
modulus of these slowly varying envelopes are shown for the first four
families (note that we have chosen $\theta=\pi/4$ for these figures).

A most important difference occurs between the two cases corresponding to $f$
odd or even: For even $f$ the classical emission $\bar{A}_{s}\left(
\mathbf{r}\right)  $ is rotationally symmetric as $\Psi_{f/2}^{0}\left(
\mathbf{r}\right)  $ is [see Eq. (\ref{modetes})], while for odd $f$ the
emitted pattern breaks that symmetry.%
\end{subequations}

\section{Squeezing}

Now we perform the analysis of quantum fluctuations affecting the classical
emission above threshold. First, in order to make the calculations as simple
as possible, we consider the limit $\gamma_{p}\gg\gamma_{s}$ in which the pump
mode can be adiabatically eliminated in Eqs. (\ref{Langevin}) as $\alpha
_{00}=\mathcal{E}_{p}/\gamma_{p}-%
%TCIMACRO{\dsum \limits_{l}}%
%BeginExpansion
{\displaystyle\sum\limits_{l}}
%EndExpansion
\frac{\chi_{l}/\gamma_{p}}{1+\delta_{0l}}\alpha_{+l}\alpha_{-l}$. The
resulting equations read
\begin{subequations}
\label{quantum eqs}%
\begin{align}
\dot{\alpha}_{\pm l}  &  =\gamma_{s}\left(  -\alpha_{\pm l}+\kappa_{l}%
N\alpha_{\mp l}^{+}\right)  +\sqrt{\gamma_{s}\kappa_{l}N}\xi_{\pm l}\left(
t\right)  ,\label{quantum eq}\\
\dot{\alpha}_{\pm l}^{+}  &  =\gamma_{s}\left(  -\alpha_{\pm l}^{+}+\kappa
_{l}N^{+}\alpha_{\mp l}^{+}\right)  +\sqrt{\gamma_{s}\kappa_{l}N^{+}}\xi_{\pm
l}^{+}\left(  t\right)  , \label{quantum eqplus}%
\end{align}
where%
\end{subequations}
\begin{equation}
N=\sigma-g^{2}%
%TCIMACRO{\dsum \limits_{l^{\prime}}}%
%BeginExpansion
{\displaystyle\sum\limits_{l^{\prime}}}
%EndExpansion
\frac{\kappa_{l^{\prime}}}{1+\delta_{0,l^{\prime}}}\alpha_{+l^{\prime}}%
\alpha_{-l^{\prime}}, \label{NL}%
\end{equation}
and $N^{+}$ is as $N$ but replacing $\alpha_{l^{\prime}}$ by $\alpha
_{l^{\prime}}^{+}$. Moreover, as we will study fluctuations in the linear
approximation, the only nonlinearities contained in $N$ that will give a
contribution correspond to the classically excited modes ($l^{\prime}=l_{0}$)
and hence%
\begin{equation}
N=\sigma-g^{2}\frac{\alpha_{+l_{0}}\alpha_{-l_{0}}}{1+\delta_{0,l_{0}}},
\label{NL1}%
\end{equation}
which is the expression we will use in the following.

We are going to distinguish between the special cases $l=l_{0}$ on one hand
and cases with $l>l_{0}$, as their properties turn out to be quite different.

\subsection{Special cases $l=l_{0}$}

Here we analyze the squeezing properties of the modes that are classically
excited. In this case $\kappa_{l}=\kappa_{l_{0}}=1$ [see Eq. (\ref{parameters}%
)], and the Langevin equations (\ref{quantum eqs}) and (\ref{NL1}) become

\begin{itemize}
\item Case $l_{0}=0$
\begin{subequations}
\label{L0}%
\begin{align}
\dot{\alpha}_{0}  &  =\gamma_{s}\left(  -\alpha_{0}+N\alpha_{0}^{+}\right)
+\sqrt{\gamma_{s}N}\xi_{0}\left(  t\right)  ,\\
N  &  =\sigma-\tfrac{1}{2}g^{2}\alpha_{0}\alpha_{0},
\end{align}

\item Case $l_{0}=1$
\end{subequations}
\begin{subequations}
\label{L1}%
\begin{align}
\dot{\alpha}_{\pm1}  &  =\gamma_{s}\left(  -\alpha_{\pm1}+N\alpha_{\pm1}%
^{+}\right)  +\sqrt{\gamma_{s}N}\xi_{\pm1}\left(  t\right)  ,\\
N  &  =\sigma-g^{2}\alpha_{+1}\alpha_{-1}%
\end{align}

\end{subequations}
\end{itemize}

We note that in neither case the information about the family order $f$
appears (up to here, that information appeared through the reduced coupling
parameters $\kappa_{l}$ [see Eqs. (\ref{parameters}), (\ref{coupling}) and
(\ref{IL})]). Moreover, Eqs. (\ref{L0}) coincide with those for the
paradigmatic single-mode DOPO (see, e.g., \cite{WallsMilburn}), while Eqs.
(\ref{L1}) do with those for the recently studied DOPO tuned to its first
transverse mode family \cite{Rotational, Notation}. The conclusions are straightforward:

\begin{itemize}
\item[(1)] For any even $f$ ($l_{0}=0$) the quantum properties of the
classically excited mode coincide with those of the usual single-mode DOPO in
the linear approximation. At threshold ($\sigma=1$) ideal squeezing is
predicted in this approximation and this squeezing degrades as we move apart
from threshold. That is why we speak about \textit{critical squeezing} in this
case, as the system parameters need to be tuned to a particular value for
obtaining the optimum squeezing level. In mathematical terms, this is a
consequence of the existence, at the bifurcation point, of a null eigenvalue
and of a companion negative eigenvalue that reaches its minimum possible
value. The former is the responsible for the "infinite"\ fluctuations of the
amplitude quadrature of the signal mode (which is proportional to the
eigenvector associated to the null eigenvalue), while the latter is the
responsible of the complete suppression (in the linear approximation) of the
fluctuations in the signal mode phase quadrature. (See below for a more
comprehensive explanation.)

\item[(2)] For any odd $f$ ($l_{0}=1$) the quantum properties of the
classically excited mode coincide again with those of the single-mode DOPO,
i.e., it shows perfect squeezing only when working exactly at the bifurcation.
Unlike the previous case the classical emission breaks the rotational
invariance of the system and then, apart from the previous critical squeezing
property, the system exhibits perfect squeezing in the remaining mode that is
spatially crossed with respect to the classically emitted one
\cite{Rotational}. As the reason for this lays on the rotational symmetry
breaking (and then on OAM conservation \cite{Rotational}), it has nothing to
do with bifurcations and the predicted squeezing is \textit{noncritical}: The
same squeezing level (perfect, in this case) is obtained at any pumping level.
From a mathematical viewpoint one of the eigenvalues governing the stability
of the classical solution above threshold is always zero (the eigenvector
associated to this eigenvalue is said to be a Goldstone mode), reflecting the
indeterminacy of the phase difference between the two signal modes [see Eq.
(\ref{stationary above})]. This phase difference is nothing but the
orientation of the emitted hybrid mode (the TEM$_{10}$ mode in
\cite{Rotational}) that results from the coherent superposition of the
$\Psi_{\left(  f-1\right)  /2}^{+1}$ and $\Psi_{\left(  f-1\right)  /2}^{-1}$
modes. Thus the physical meaning of the Goldstone mode is that the orientation
of the signal mode emitted above threshold by this DOPO\ diffuses with time,
and is thus undetermined. This eigenvalue is accompanied by another eigenvalue
which always takes its minimum possible value, with the consequence that the
amplitude of its associated eigenvector has no fluctuations at all. This last
eigenvector can be easily identified with the OAM of the emitted pattern,
which is nothing but another hybrid mode spatially crossed (orthogonal)\ with
respect to the bright one. This shows that the generation of noncritical
squeezing through the spontaneous rotational symmetry breaking mechanism is
not particular of the model considered in \cite{Rotational}, but should be a
general phenomenon. We address the reader to \cite{Rotational} for full
details of this phenomenon.
\end{itemize}

\subsection{Cases $l>l_{0}$: Squeezing of higher OAM modes}

In this case the amplitudes $\alpha_{l}$ and $\alpha_{l}^{+}$ are very small
(they are null in the classical limit even if the system is above threshold)
and then, in the linear approximation we are considering, the nonlinear
function $N$ (\ref{NL1}) appearing in (\ref{quantum eqs}) must be evaluated at
its classical value (\ref{stationary above}) as not doing would introduce
higher order corrections that must be neglected in the used linear
approximation. One then obtains $N=1$ and the Langevin equations become
\begin{subequations}
\label{quantum higher}%
\begin{align}
\dot{\alpha}_{\pm l}  &  =\gamma_{s}\left(  -\alpha_{\pm l}+\kappa_{l}%
\alpha_{\mp l}^{+}\right)  +\sqrt{\gamma_{s}\kappa_{l}}\xi_{\pm l}\left(
t\right)  ,\\
\dot{\alpha}_{\pm l}^{+}  &  =\gamma_{s}\left(  -\alpha_{\pm l}^{+}+\kappa
_{l}\alpha_{\mp l}^{+}\right)  +\sqrt{\gamma_{s}\kappa_{l}}\xi_{\pm l}%
^{+}\left(  t\right)  .
\end{align}
As is evident, the evolution of the different pairs of OAM $\pm l$ is
decoupled, what allows us to analyze separately the quantum properties of each
$l$ couple. Moreover, the evolution of these fluctuations does not depend on
whether the family is odd or even.

The study of Eqs. (\ref{quantum higher}) is facilitated by expressing them in
vector form as%

\end{subequations}
\begin{equation}
\mathbf{\dot{a}}_{l}=\mathcal{L}_{l}\mathbf{a}_{l}+\sqrt{\gamma_{s}\kappa_{l}%
}\boldsymbol{\xi}_{l}\left(  t\right)  , \label{MatLinLan-l}%
\end{equation}
with%
\begin{subequations}
\begin{align}
\mathbf{a}_{l}  &  =\operatorname{col}\left(  \alpha_{+l},\alpha_{+l}%
^{+},\alpha_{-l},\alpha_{-l}^{+}\right)  ,\\
\boldsymbol{\xi}_{l}  &  =\operatorname{col}\left(  \xi_{l}\left(  t\right)
,\xi_{l}^{+}\left(  t\right)  ,\xi_{l}^{\ast}\left(  t\right)  ,\left[
\xi_{l}^{+}\left(  t\right)  \right]  ^{\ast}\right)  ,\\
\mathcal{L}_{l}  &  =\gamma_{s}%
\begin{pmatrix}
-1 & 0 & 0 & \kappa_{l}\\
0 & -1 & \kappa_{l} & 0\\
0 & \kappa_{l} & -1 & 0\\
\kappa_{l} & 0 & 0 & -1
\end{pmatrix}
,
\end{align}
where we remind that $0<\kappa_{l}<1$ [see Eqs. (\ref{parameters})].

Linear matrix $\mathcal{L}_{l}$ has eigenvalues $\lambda_{1}^{l}=\lambda
_{2}^{l}=-\gamma_{s}\left(  1-\kappa_{l}\right)  $ and $\lambda_{3}%
^{l}=\lambda_{4}^{l}=-\gamma_{s}\left(  1+\kappa_{l}\right)  $ with
corresponding eigenvectors \cite{nota}
\end{subequations}
\begin{subequations}
\label{eigen}%
\begin{align}
\mathbf{w}_{1,2}^{l}  &  =\frac{1}{2}\operatorname{col}\left(  1,\pm
1,\pm1,1\right)  ,\\
\mathbf{w}_{3,4}^{l}  &  =\frac{1}{2}\operatorname{col}\left(  1,\pm
1,\mp1,-1\right)  .
\end{align}
By projecting the linear Langevin system (\ref{MatLinLan-l}) onto
$\mathbf{w}_{j}^{l}$ we find
\end{subequations}
\begin{subequations}
\label{c's}%
\begin{align}
\dot{c}_{1,4}^{l}  &  =-\gamma_{s}\left(  1\mp\kappa_{l}\right)  c_{1}%
^{l}+\sqrt{\gamma_{s}\kappa_{l}}\eta_{1,4}^{l}\left(  t\right)  ,\\
\dot{c}_{2,3}^{l}  &  =-\gamma_{s}\left(  1\mp\kappa_{l}\right)  c_{3}%
^{l}+i\sqrt{\gamma_{s}\kappa_{l}}\eta_{2,3}^{l}\left(  t\right)  ,
\end{align}
where we defined the projections%
\end{subequations}
\begin{equation}
c_{j}^{l}\left(  t\right)  =\mathbf{w}_{j}^{l}\cdot\mathbf{a}_{l}\left(
t\right)  ,
\end{equation}
and the four new real noises%
\begin{subequations}
\begin{align}
\eta_{1,4}^{l}\left(  t\right)   &  =\mathbf{w}_{1,4}^{l}\cdot\mathbf{\xi}%
_{l}\left(  t\right)  =\operatorname{Re}\left[  \xi_{l}\left(  t\right)
\pm\xi_{l}^{+}\left(  t\right)  \right]  ,\\
\eta_{2,3}^{l}\left(  t\right)   &  =-i\mathbf{w}_{2,3}^{l}\cdot\mathbf{\xi
}_{l}\left(  t\right)  =\operatorname{Im}\left[  \xi_{l}\left(  t\right)
\mp\xi_{l}^{+}\left(  t\right)  \right]  ,
\end{align}
which satisfy the statistical properties%
\end{subequations}
\begin{equation}
\left\langle \eta_{j}^{l}\left(  t\right)  \right\rangle =0,\ \ \left\langle
\eta_{m}^{l}\left(  t_{1}\right)  \eta_{n}^{l}\left(  t_{2}\right)
\right\rangle =\delta_{mn}\delta\left(  t_{1}-t_{2}\right)  . \label{r-corr}%
\end{equation}

Equations (\ref{c's}) are solved with standard methods. The variance spectra
for projections $c_{j}$, defined as%
\begin{equation}
\tilde{C}_{j}\left(  \omega\right)  =\int_{-\infty}^{+\infty}d\tau
e^{-i\omega\tau}\left\langle c_{j}\left(  t\right)  ,c_{j}\left(
t+\tau\right)  \right\rangle ,
\end{equation}
read, in the stationary limit $t\gg\lambda_{j}^{-1}$,%
\begin{equation}
\tilde{C}_{1,3}^{l}\left(  \omega\right)  =-\tilde{C}_{2,4}^{l}\left(
\omega\right)  =\pm\frac{\gamma_{s}\kappa_{l}}{\left[  \gamma_{s}\left(
1\mp\kappa_{l}\right)  \right]  ^{2}+\omega^{2}}.
\end{equation}

The advantage of the eigensystem method we are using
\cite{EPL,translationalPRA,Rotational} is that the relevant quadratures of the
problem appear in a natural way, as the projections $c_{j}^{l}$ are related to
the problem quadratures through%
\begin{subequations}
\begin{align}
X_{\mathrm{c},l}  &  =\sqrt{2}c_{1}^{l},\ \ X_{\mathrm{s},l}=i\sqrt{2}%
c_{2}^{l},\\
Y_{\mathrm{c},l}  &  =-i\sqrt{2}c_{4}^{l},\ \ Y_{\mathrm{s},l}=\sqrt{2}%
c_{3}^{l},
\end{align}
where%
\end{subequations}
\begin{align}
X_{j,l}  &  =\left(  \alpha_{j,l}^{+}+\alpha_{j,l}\right)  ,\\
Y_{j,l}  &  =i\left(  \alpha_{j,l}^{+}+\alpha_{j,l}\right)  ,
\end{align}
refer to the hybrid mode $H_{j,\left(  f-l\right)  /2}^{l}\left(
\mathbf{r}\right)  $ as it can be easily seen from Eqs. (\ref{GLtoH}). These
simple relations are the consequence of the appropriate choice of the
eigenvectors of $\mathcal{L}_{l}$ \cite{nota}.

We are now in conditions to find out the squeezing spectra of the quadratures.
By using Eq. (\ref{Sout}) together with the relations above, we get that
$S_{x,\mathrm{c},l}^{\mathrm{out}}\left(  \omega\right)  =S_{x,\mathrm{s}%
,l}^{\mathrm{out}}\left(  \omega\right)  =4\gamma_{s}\tilde{C}_{1}^{l}\left(
\omega\right)  $ and $S_{y,\mathrm{c},l}^{\mathrm{out}}\left(  \omega\right)
=S_{y,\mathrm{s},l}^{\mathrm{out}}\left(  \omega\right)  =4\gamma_{s}\tilde
{C}_{3}^{l}\left(  \omega\right)  $, i.e., the squeezing of the phase
quadratures are%
\begin{equation}
S_{y,j,l}^{\mathrm{out}}\left(  \omega\right)  =-\frac{4\kappa_{l}}{\left(
1+\kappa_{l}\right)  ^{2}+\left(  \omega/\gamma_{s}\right)  ^{2}%
},\ \ j=\mathrm{c},\mathrm{s}. \label{result}%
\end{equation}
This expression clearly shows that the phase quadratures $Y_{j,l}$
($j=\mathrm{c},\mathrm{s}$) of both $H_{\mathrm{c},\left(  f-l\right)  /2}%
^{l}\left(  \mathbf{r}\right)  $ and $H_{\mathrm{s},\left(  f-l\right)
/2}^{l}\left(  \mathbf{r}\right)  $ hybrid modes (i) have the same
fluctuations properties, (ii) are noncritically squeezed as $S_{y,j,l}%
^{\mathrm{out}}\left(  \omega\right)  <0$ at any noise frequency $\omega$ and
this value is independent of the distance from threshold [it is however
dependent on the OAM value through the value of the ratio $\kappa_{l}$
(\ref{parameters})], (iii) and exhibit maximum squeezing at $\omega=0$. This
is a main results of our article and we pass to discuss it.

\begin{figure}[ht]

\includegraphics[
width=3.3in
]%
{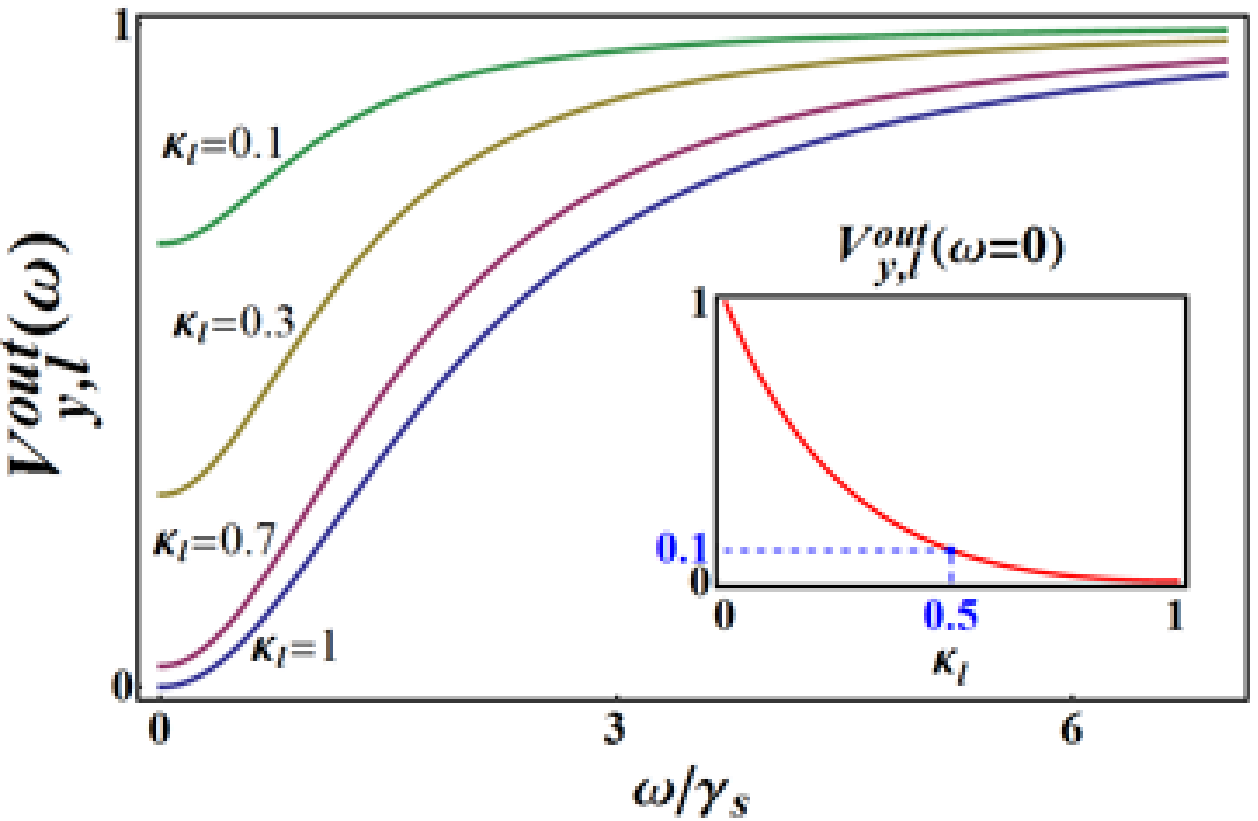}%
\\
{\small Figure 2.- Variance spectra of the phase quadratures of some
Hybrid modes with $l\neq l_0$. The inset
shows the same quantity at $\omega=0$ as a
function of $\kappa_l$.}%

\end{figure}

\subsection{Discussion}

We have shown that the light emitted by a DOPO with a large-Fresnel-number
cavity with spherical mirrors exhibits a variety of squeezing properties. On
the one hand there is the usual critical squeezing appearing at the
oscillation threshold \cite{WallsMilburn} associated with the classically
emitted solution. There is also the noncritical squeezing due to the
spontaneous breaking of the rotational symmetry that appears above threshold
for odd $f$ ($l_{0}=1$) \cite{Rotational}. Also a new type of squeezing
behavior has been derived for all hybrid modes with $l>l_{0}$, which remain
off at the classical level once the oscillation threshold is crossed,
consisting in that their phase quadratures are noncritically squeezed for any
$l$, as shown by Eq. (\ref{result}).%

The variance spectrum for either of the phase quadratures $V_{y,l}%
^{\mathrm{out}}\left(  \omega\right)  =1+S_{y,j,l}^{\mathrm{out}}\left(
\omega\right)  $ for $j=\mathrm{c}$ or $j=\mathrm{s}$, is represented in Fig.
2 for different values of $\kappa_{l}$ [see Eq. (\ref{parameters})]. In the
figure, $V_{y,l}^{\mathrm{out}}\left(  \omega=0\right)  $ is also shown, at
the inset, as a function of $\kappa_{l}$. Clearly, as $\kappa_{l}=\chi
_{l}/\chi_{l_{0}}\rightarrow1$, what occurs as $l\rightarrow l_{0}$, nearly
perfect squeezing is found at $\omega=0$. Notice that more than $90\%$ of
squeezing can be achieved if $\kappa_{l}>0.5$. But the ratio $\kappa_{l}%
=I_{l}/I_{l_{0}}$, which is obtained from Eq. (\ref{coupling}), is solely
determined by geometrical reasons, and thus the squeezing properties of these
higher OAM modes do not depend on the system parameters (of course we must
remind that we assumed equal cavity losses for all modes).

In order to better appreciate the large amounts of squeezing exhibited by
these modes, we give their associated noise reduction in percentage for
different even (odd) families in Table 1 (2). Notice that the largest
squeezing occurs for large values of $f$ and small values of $l$, and works
better for even families.%

\begin{figure}[ht]

\includegraphics[
width=3.3in
]%
{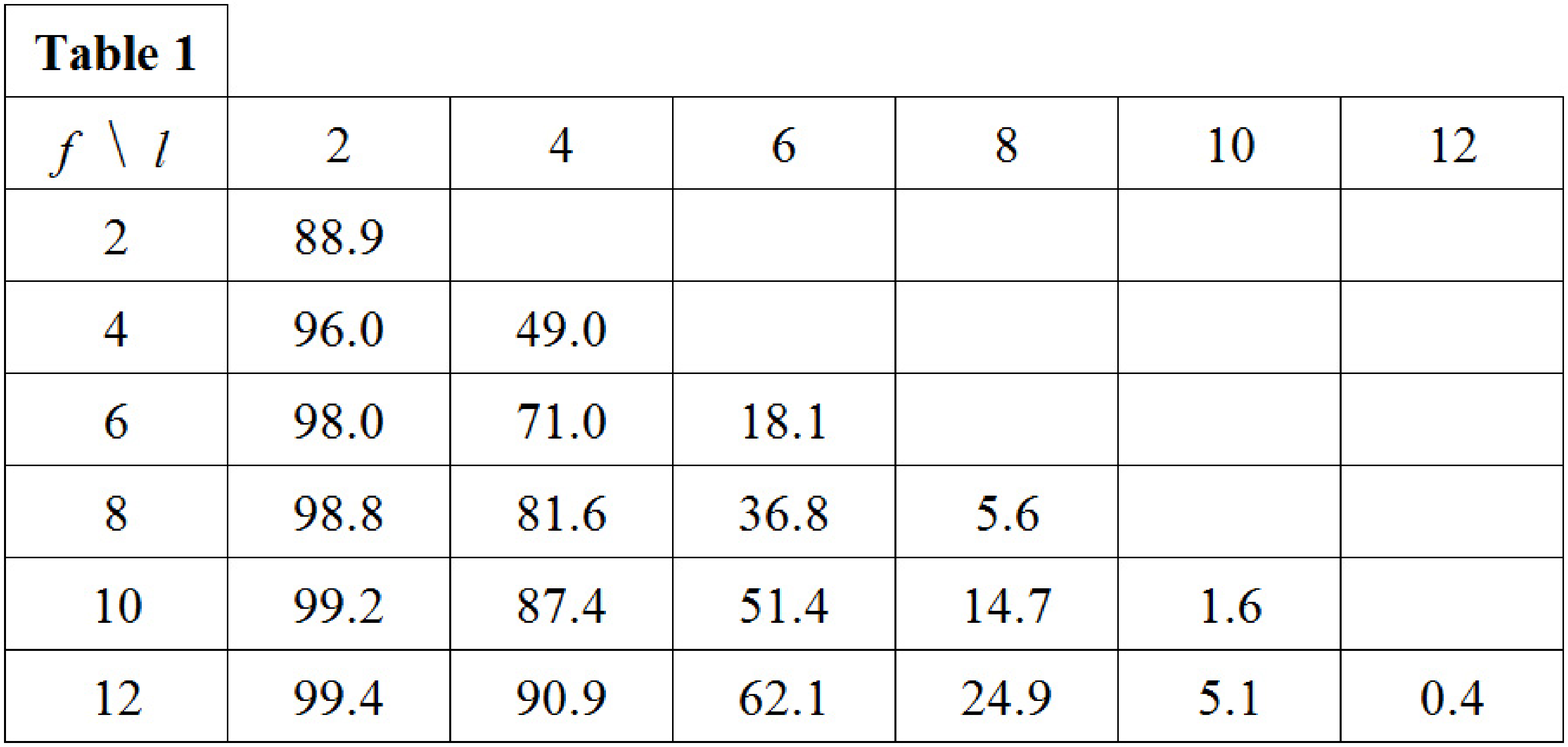}%
\\
{\small Table 1.- Percentage of noise reduction for the non amplified
Hybrid modes $l\neq l_0$ laying in even families. Note that
large levels of squeezing are obtained for the lower angular momentum modes.}%

\end{figure}

\begin{figure}[ht]

\includegraphics[
width=3.3in
]%
{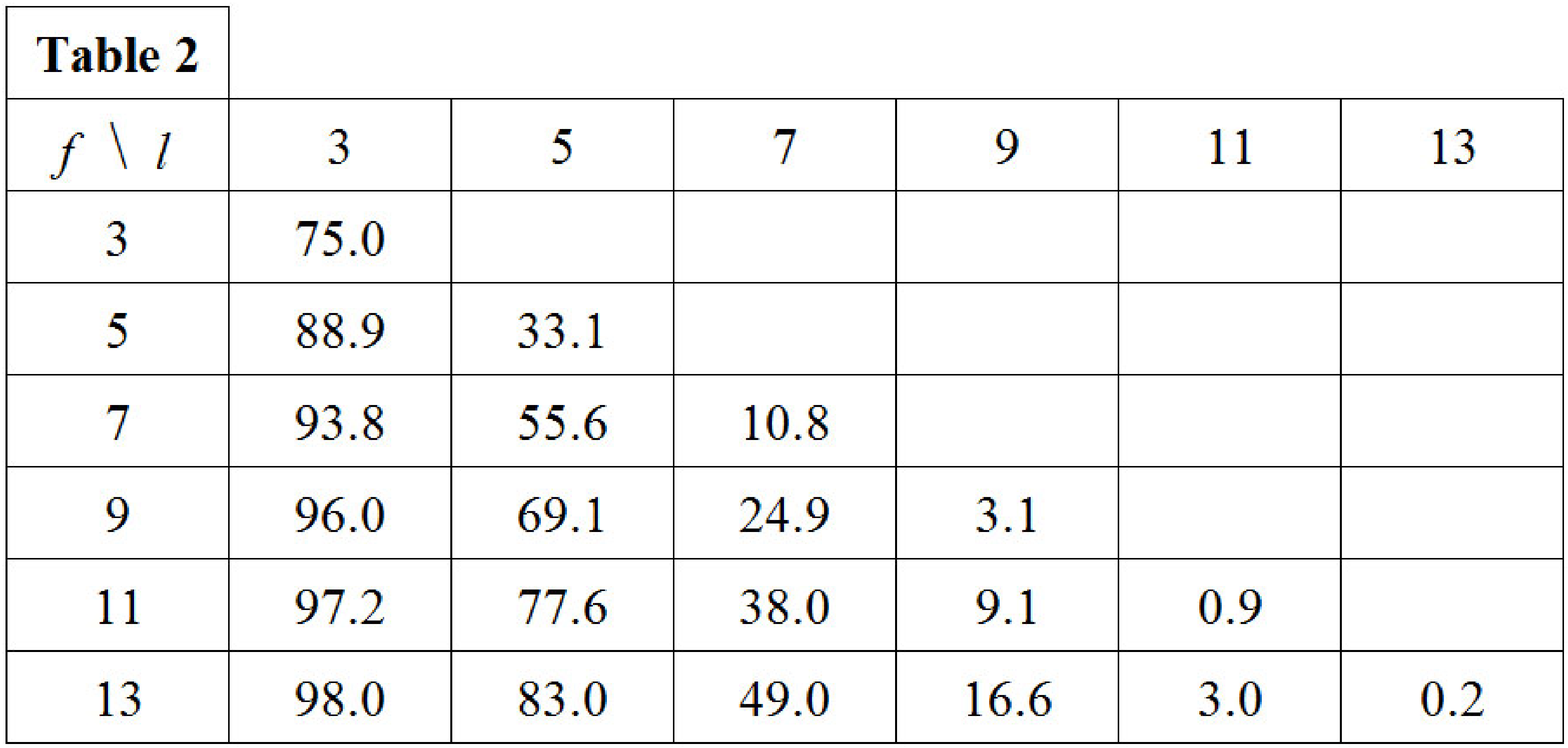}%
\\
{\small Table 2.- Same as Table 1, but for odd families. Large levels
of squeezing are also obtained in this case for the lower angular momentum
modes.}%

\end{figure}

It is to be remarked that both the hybrid mode $H_{\mathrm{c},n}^{l}\left(
\mathbf{r}\right)  $ and its orthogonal $H_{\mathrm{s},n}^{l}\left(
\mathbf{r}\right)  $ have the same squeezing properties. This means that the
orientation of the mode is irrelevant, as an hybrid mode rotated an arbitrary
angle $\beta$ respect to the $x$ axis, which is given by%
\begin{align}
H_{\beta,n}^{l}\left(  \mathbf{r}\right)   &  =H_{\mathrm{c},n}^{l}\left(
r,\phi-\beta\right) \\
&  =H_{\mathrm{c},n}^{l}\left(  \mathbf{r}\right)  \cos\left(  l\beta\right)
+H_{\mathrm{s},n}^{l}\left(  \mathbf{r}\right)  \sin\left(  l\beta\right)  ,
\end{align}
also has the same squeezing properties. This is in clear contrast to the
perfectly squeezed mode in the $l=l_{0}=1$ case \cite{Rotational}, which has
not an arbitrary orientation but is orthogonal to the classically excited mode
at every instant. Of course, the reason is that the modes here analyzed are
below their threshold, at difference with the case in \cite{Rotational}.

So far we have proven that the higher OAM modes show noise reduction in their
phase quadrature. Let us see why is this occurring. The coefficient that rules
the squeezing properties of the $l\neq l_{0}$ modes is $\kappa_{l}$, see Eq.
(\ref{result}). Both above and below threshold, $\kappa_{l}$ can be written in
terms of the stationary pump field as $\kappa_{l}=\chi_{l}\bar{\alpha}%
_{00}/\gamma_{s}$, see (\ref{parameters}). Below threshold $\bar{\alpha}_{00}$
increases linearly with the external pump amplitude [see Eq.
(\ref{stationary below})] and hence so does $\kappa_{l}$. But once the
bifurcation is reached and the DOPO starts emitting the $l_{0}$ mode(s), the
value of $\bar{\alpha}_{00}$ saturates, and remains constant irrespective of
the pump value [see Eq. (\ref{stationary above})] leading to $\kappa_{l}%
=I_{l}/I_{l_{0}}$ above threshold, which is independent of the pump value.
This implies that the value of $\kappa_{l}$ will remain fixed to that at the
bifurcation once the DOPO is above threshold and, consequently, the squeezing
level of these modes becomes noncritical.

\section{Conclusions}

We have shown that tuning large Fresnel number DOPOs with spherical mirrors to
transverse families is a simple way for generating squeezed light with the
shape of hybrid Laguerre--Gauss modes with OAM $l>l_{0}$, with $l_{0}=0$ or
$1$ depending on the even or odd character of the selected transverse mode
family. The behavior of the system can be resumed as follows: Above
threshold, only the hybrid mode with $l=l_{0}$ is amplified, and the rest of
modes (those with $l>l_{0}$) remain off. The amplified modes exhibit squeezing
properties that can be found in the literature (see \cite{WallsMilburn} for
$l_{0}=0$ and \cite{Rotational} for $l_{0}=1$) as shown in Section IV A.

The surprising result is that hybrid modes with $l>l_{0}$ exhibit a large
degree of squeezing together with the fact that this squeezing is noncritical,
i.e., it remains fixed irrespective of the pump level when the system is above
threshold. Thus, squeezed vacua with the shape of higher order hybrid modes
can be generated by normal DOPOs if these have a large Fresnel number.

In our model we assumed equal cavity losses for all signal modes (in fact,
equal cavity losses for all signal modes belonging to the same family; this is
the precise meaning we give to the expression large Fresnel number cavity). As
we have explained, the squeezing level of modes with OAM $l>l_{0}$ depends on
the distance between their threshold and that for the mode(s) with lowest OAM
$l_{0}$. Hence, differences in the values of their decay rates could modify
quantitatively the results we have presented, as this would change the
threshold for the different modes. If fact, if tailoring the cavity losses for
the different modes would be possible, one could obtain even smaller quantum
fluctuations for these empty modes if their corresponding thresholds were made
closer to that of the $l_{0}$ mode. We also assumed perfect cavity resonance
for a particular family of modes. We do not expect that detuning changes our
main conclusions, as detuning affects equally all modes within the same family
(apart, of course, from the fact that if the detuning is half the value the
transverse free spectral range, competition phenomena between the involved
families will manifest, something that is not considered in our model). Adding
detuning would have the same consequences as in the single-mode DOPO: It will
increase the threshold and change the phase of the amplified mode (and hence
of the squeezed quadratures).

We finally note that Laguerre--Gauss modes are becoming important for several
purposes, e.g. in quantum information processing and in manipulation of atoms
\cite{Arnold08} just to mention a couple. We hope that our results can be of
relevance as squeezed hybrid modes or, equivalently, entangled Laguerre-Gauss
modes, can be of utility for these purposes.

This work has been supported by the Spanish Ministerio de Eduaci\'{o}n y
Ciencia and the European Union FEDER through Projects FIS2005-07931-C03-01 and
FIS2008-06024-C03-01. C.N.-B. is a grant holder of the FPU program of the
Ministerio de Educaci\'{o}n y Ciencia (Spain).

\appendix

\section{Cavity modes}

For the sake of clarity, we find it convenient to review the main properties
of the cavity modes in a Fabry-Perot resonator with spherical mirrors (see,
e.g., \cite{Hodgson} for more details). Within the paraxial approximation, it
is well known that the Laguerre-Gauss modes form a complete set of spatial
modes describing the light inside the resonator. Let $R_{1}$ and $R_{2}$
denote the curvature radius of the cavity mirrors, and $L_{\mathrm{eff}%
}=L-\left(  1-1/n_{c}\right)  l_{c}$ the effective cavity length, being $L$
the geometrical length of the resonator, and $l_{c}$ and $n_{c}$ the length
and refractive index, respectively, of the $\chi^{\left(  2\right)  }$
crystal. Then the Laguerre-Gauss modes at the resonator waist plane can be
written as%
\begin{equation}
\Psi_{n}^{\pm l}\left(  \mathbf{r}\right)  =\mathcal{N}_{n}^{l}u_{n}%
^{l}\left(  r\right)  \exp\left(  \pm il\phi\right)  , \label{GL}%
\end{equation}
with $\mathbf{r}=\left(  x,y\right)  $ the transverse coordinates being
$\mathbf{r}=r\left(  \cos\phi,\sin\phi\right)  $ its polar decomposition,
$\mathcal{N}_{n}^{l}$ a normalization factor and%
\begin{equation}
u_{n}^{l}\left(  r\right)  =\frac{1}{w}\left(  \frac{\sqrt{2}r}{w}\right)
^{l}L_{n}^{l}\left(  \frac{2r^{2}}{w^{2}}\right)  e^{-r^{2}/w^{2}},
\end{equation}
being $L_{n}^{l}$ the modified Laguerre polynomial with radial and polar
indices $n,l\in%
%TCIMACRO{\U{2115} }%
%BeginExpansion
\mathbb{N}
%EndExpansion
$, which are given by Rodrigues formula%
\begin{equation}
L_{n}^{l}\left(  v\right)  =\frac{1}{n!}e^{v}\frac{1}{v^{l}}\frac{d^{n}%
}{dv^{n}}\left(  e^{-v}v^{l}v^{n}\right)  .
\end{equation}
By choosing the normalization factor as%
\begin{equation}
\mathcal{N}_{n}^{l}=\sqrt{\frac{2}{\pi}\frac{n!}{\left(  n+l\right)  !}},
\end{equation}
the following orthogonality relation holds%
\begin{equation}
\int_{0}^{2\pi}d\phi\int_{0}^{\infty}rdr\left[  \Psi_{n}^{\pm l}\left(
\mathbf{r}\right)  \right]  ^{\ast}\Psi_{n^{\prime}}^{\pm l^{\prime}}\left(
\mathbf{r}\right)  =\delta_{ll^{\prime}}\delta_{nn^{\prime}}.
\end{equation}

The beam spot size at the cavity waist, $w$, is given by%
\begin{equation}
w^{2}=\frac{2cL_{\mathrm{eff}}}{\omega}\frac{\sqrt{g_{1}g_{2}\left(
1-g_{1}g_{2}\right)  }}{g_{1}+g_{2}-2g_{1}g_{2}}, \label{w}%
\end{equation}
with $g_{j}=1-L_{\mathrm{eff}}/R_{j}$ and $\omega$ the beam frequency.

The Laguerre-Gauss basis is recommended in order to visualize the OAM of the
field, as these modes are eigenstates of the OAM operator $-i\partial_{\phi}$
with eigenvalues $\pm l$. Concerning the resonance frequency of the different
$\Psi_{n}^{\pm l}$ modes, they are different in general for each mode.
Concretely $\omega_{qnl}=q\pi c/L_{\mathrm{eff}}+\Delta\omega_{nl}$, with $q$
an integer (different $q^{\prime}s$ correspond to different longitudinal
cavity modes) and the transverse part of the resonance frequency is given by%
\begin{equation}
\Delta\omega_{nl}=\frac{c}{L_{\mathrm{eff}}}\left(  1+2n+l\right)
\arccos\left(  \sqrt{g_{1}g_{2}}\right)  .
\end{equation}

\begin{figure}[t]

\includegraphics[
width=3.3in
]%
{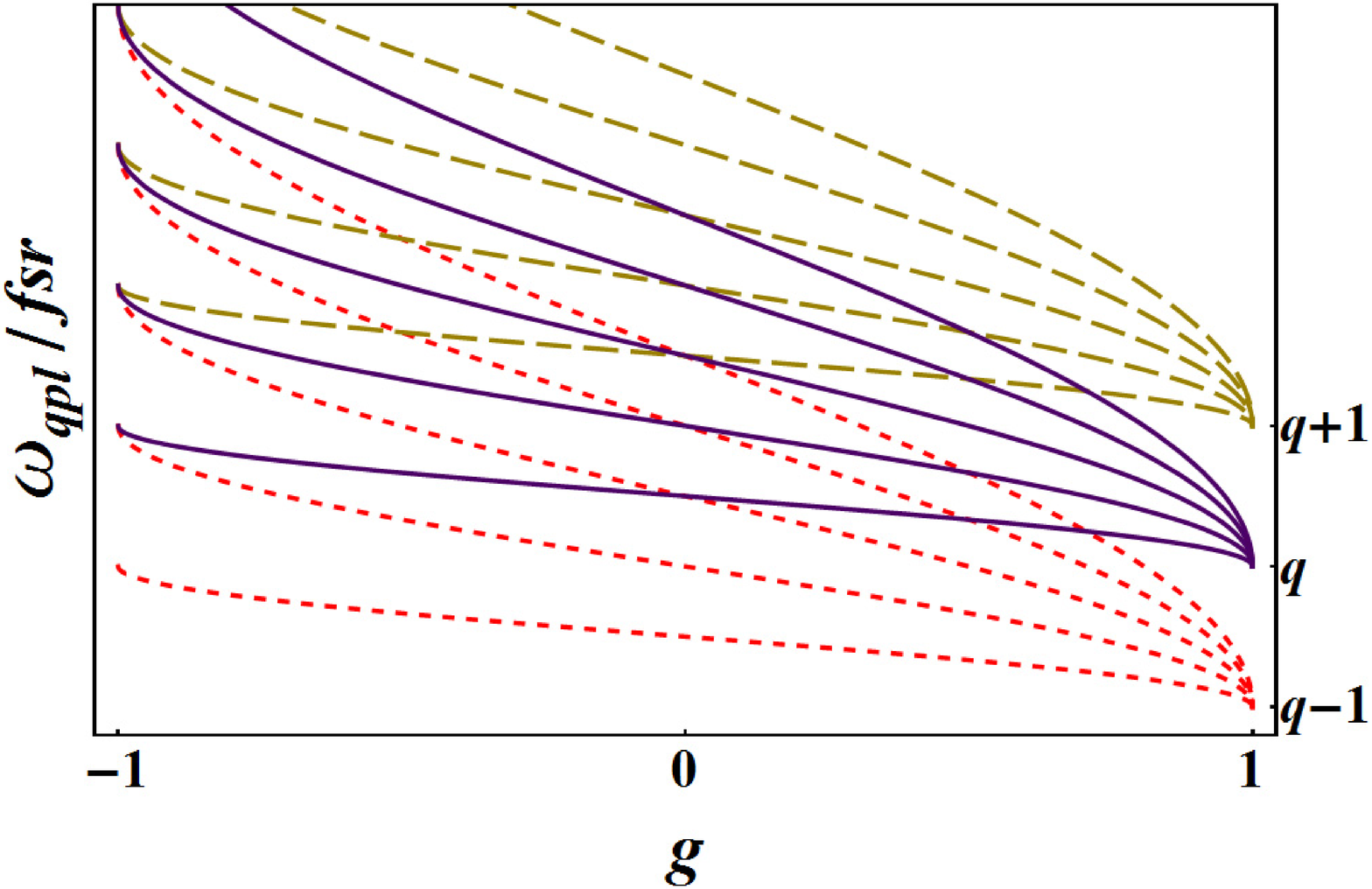}%
\\
{\small Figure 3.- Resonance frequencies of the first 4 families
corresponding to three consecutive longitudinal modes, as a function of the $g$ parameter of a symmetrical resonator. Different families corresponding to the same longitudinal mode have the same color and dashing.
The family order $f$ increases with the frequency.}%

\end{figure}

Hence, cavity modes having the same \textit{family order} $f=\left(
2n+l\right)  $ have\ the same frequency and are said to be members of the same
family $f$. It is clear that family $f$ consists of the set of $f+1$
Laguerre-Gauss modes $\left\{  \Psi_{\left(  f-l\right)  /2}^{\pm l}\right\}
$, with $l=f,f-2,...,l_{0}$, having OAM $\pm f$, $\pm\left(  f-2\right)
$,...,$\pm l_{0}$, respectively. The lower OAM modes have $l_{0}=0$ or $1$ for
even or odd $f$ respectively. In Fig. 3 we represent $\omega_{qnl}$ (in units
of the free spectral range, $fsr=\pi c/L_{\mathrm{eff}}$) for several modes as
a function of $g$ for \textit{symmetric} resonators (those for which
$g_{1}=g_{2}\equiv g$) which are \textit{stable} (i.e., $-1<g<1$; be reminded
that $g=1$ corresponds to a planar mirrors resonator, $g=0$ to a confocal
resonator, and $g=-1$ to a concentric resonator). It can be clearly
appreciated that different transverse families corresponding to different
longitudinal modes can have the same frequency for rational values of $\left(
\arccos g\right)  /\pi$. Below we assume that different families have
different frequencies, i.e., that this quantity has an irrational value.%

\section{Langevin equations}

Next we outline the derivation of the Langevin Eqs. (\ref{Langevin}).
Neglecting the effect of thermal photons, the master equation ruling the
evolution of the system's density operator \cite{Carmichael} reads
\begin{align}
\frac{\partial}{\partial t}\hat{\varrho}  &  =\frac{1}{i\hbar}\left[  \hat
{H},\hat{\varrho}\right] \nonumber\\
&  +%
%TCIMACRO{\dsum \limits_{\pm l}}%
%BeginExpansion
{\displaystyle\sum\limits_{\pm l}}
%EndExpansion
\gamma_{s}\left(  2\hat{a}_{l}\hat{\varrho}\hat{a}_{l}^{\dagger}-\hat{a}%
_{l}^{\dagger}\hat{a}_{l}\hat{\varrho}-\hat{\varrho}\hat{a}_{l}^{\dagger}%
\hat{a}_{l}\right) \nonumber\\
&  +\gamma_{p}\left(  2\hat{a}_{00}\hat{\varrho}\hat{a}_{00}^{\dagger}-\hat
{a}_{00}^{\dagger}\hat{a}_{00}\hat{\varrho}-\hat{\varrho}\hat{a}_{00}%
^{\dagger}\hat{a}_{00}\right)  \label{Master}%
\end{align}
where losses are assumed to occur in only one of the cavity mirrors and
$\hat{H}$ is given by Eq. (\ref{Hamiltonian}). Now we use the positive $P$
representation \cite{Drummond80,Gardiner00} of the density operator, which
allows the evaluation of expected values of normally ordered operators as
averages of functions in phase space. By using standard methods
\cite{Drummond80,Gardiner00}, the master equation (\ref{Master}) is then
transformed into a Fokker-Planck type equation for $P\left(
\boldsymbol{\alpha},\boldsymbol{\alpha}^{+}\right)  $, where all the coherent
amplitudes are collected into the vector%
\begin{equation}
\boldsymbol{\alpha}=\operatorname{col}\left(  \alpha_{00},\alpha_{00}%
^{+},\alpha_{+l},\alpha_{+l}^{+},\alpha_{-l},\alpha_{-l}^{+}\right)  .
\end{equation}
The equation for $P$ reads%
\begin{equation}
\frac{\partial}{\partial t}P=\left[  -\sum_{i}\frac{\partial}{\partial
\alpha_{i}}A_{i}+\frac{1}{2}\sum_{i,j}\frac{\partial^{2}}{\partial\alpha
_{i}\partial\alpha_{j}}\mathcal{D}_{ij}\right]  P,
\end{equation}
where $\alpha_{i}$ means the $i$-th element of $\boldsymbol{\alpha}$, with%
\begin{gather}
A_{1}=\mathcal{E}_{p}\mathcal{-}\gamma_{p}\alpha_{00}-%
%TCIMACRO{\dsum \limits_{l}}%
%BeginExpansion
{\displaystyle\sum\limits_{l}}
%EndExpansion
\frac{\chi_{l}}{1+\delta_{0l}}\alpha_{-l}\alpha_{+l},\\
A_{3}=-\gamma_{s}\alpha_{+l}+\chi_{l}\alpha_{00}\alpha_{+l}^{+},\\
A_{5}=-\gamma_{s}\alpha_{-l}+\chi_{l}\alpha_{00}\alpha_{-l}^{+},\\
\mathcal{D}_{35}=\mathcal{D}_{53}=\chi_{l}\alpha_{00},\text{ }\mathcal{D}%
_{46}=\mathcal{D}_{64}=\chi_{l}\alpha_{00}^{+},
\end{gather}
and $A_{2,4,6}$ are obtained from $A_{1,3,5}$ by changing $\alpha
_{j}\longleftrightarrow\alpha_{j}^{+}$. Any other element of the diffusion
matrix $\mathcal{D}$ is null.

Next, as stated by the It\^{o} theorem \cite{Arnold}, this Fokker-Planck
equation is mapped onto the set of coupled stochastic (Langevin) equations%
\begin{equation}
\boldsymbol{\dot{\alpha}}=\mathbf{A}\left(  \boldsymbol{\alpha}\right)
+\mathcal{B}\left(  \boldsymbol{\alpha}\right)  \cdot\boldsymbol{\eta}\left(
t\right)  , \label{eq-Langevin}%
\end{equation}
where $\boldsymbol{\eta}$ is a vector with real white noises as components,
each of them satisfying the statistical properties defined in Eq.
(\ref{r-corr}), $\mathbf{A}$ is a vector with components $A_{i}$ defined
above; and the \textit{noise matrix} $\mathcal{B}$ is defined by
$\mathcal{D}=\mathcal{B}\cdot\mathcal{B}^{T}$. The equivalence between the
Fokker-Planck equation and the Langevin system has to be understood as
$\left\langle f\left(  \boldsymbol{\alpha}\right)  \right\rangle
_{P}=\left\langle f\left(  \boldsymbol{\alpha}\right)  \right\rangle
_{\mathrm{stochastic}}$, i.e., phase space averages are equal to averages made
by using the statistical properties of the noise vector $\boldsymbol{\eta
}\left(  t\right)  $.

As for the diffusion matrix $\mathcal{D}$, it is composed of uncoupled minors
for each one of the $\pm l$ and $00$ subspaces, and then so does matrix
$\mathcal{B}$. In addition, the noise matrix minor associated to the pump mode
is the zero matrix as this is its associated minor in $\mathcal{D}$. One
possible choice for the minors associated to $\pm l$ modes is%
\begin{equation}
\mathcal{B}_{l}=\sqrt{\frac{\chi_{l}}{2}}%
\begin{pmatrix}
\sqrt{\alpha_{00}} & 0 & i\sqrt{\alpha_{00}} & 0\\
0 & \sqrt{\alpha_{00}^{+}} & 0 & i\sqrt{\alpha_{00}^{+}}\\
\sqrt{\alpha_{00}} & 0 & -i\sqrt{\alpha_{00}} & 0\\
0 & \sqrt{\alpha_{00}^{+}} & 0 & -i\sqrt{\alpha_{00}^{+}}%
\end{pmatrix}
,
\end{equation}
for $l\neq0$ and%
\begin{equation}
\mathcal{B}_{0}=\sqrt{\chi_{0}}%
\begin{pmatrix}
\sqrt{\alpha_{00}} & 0\\
0 & \sqrt{\alpha_{00}^{+}}%
\end{pmatrix}
,
\end{equation}
for $l=0$. Finally, inserting this noise matrix into the equivalent Langevin
system (\ref{eq-Langevin}), we get Eqs. (\ref{Langevin}).


\begin{thebibliography}{99}

                                                                                              %
\bibitem {Loudon87}R. Loudon and P. L. Knight, J. Mod. Opt. \textbf{34}, 709 (1987).

\bibitem {WallsMilburn}D. F. Walls and G. J. Milburn, \textit{Quantum Optics}
(Springer, 1994).

\bibitem {Drummond04}P. D. Drummond and Z. Ficek (eds.), \textit{Quantum
Squeezing} (Springer, 2004).

\bibitem {Gea}J. Gea--Banacloche et al., Phys. Rev. A \textbf{41}, 369 (1990).

\bibitem {Vahlbruch08}H. Vahlbruch et al., Phys. Rev. Lett. \textbf{100},
033602 (2008); see also Y. Takeno et al., Opt. Express \textbf{15}, 4321 (2007).

\bibitem {Braunstein05}S. L. Braunstein and P. van Loock, Rev. Mod. Phys.
\textbf{77}, 513 (2005).

\bibitem {Fabre}C. Fabre et al., Opt. Lett. \textbf{25}, 76 (2000); N. Treps
et al., Phys. Rev. Lett. \textbf{88}, 203601 (2002); N. Treps et al., Science
\textbf{301}, 940 (2003).

\bibitem {Arnold08}See, e.g., S. Franke--Arnold and A. S. Arnold, Am.
Sci. \textbf{96}, 226 (2008).

\bibitem {Gatti}A. Gatti, L.A. Lugiato, Phys. Rev. A \textbf{52}, 1675 (1995).

\bibitem {EPL}I. P\'{e}rez-Arjona, E. Rold\'{a}n, and G. J. de Valc\'{a}rcel,
Europhys. Lett. \textbf{74}, 247 (2006).

\bibitem {translationalPRA}I. P\'{e}rez-Arjona, E. Rold\'{a}n, and G. J. de
Valc\'{a}rcel, Phys. Rev. A \textbf{75}, 063802 (2007).

\bibitem {Petsas}K.I. Petsas, A. Gatti and L.A. Lugiato, Quantum Semiclass.
Opt. \textbf{10}, 789 (1998).

\bibitem {Rotational}C. Navarrete--Benlloch, E. Rold\'{a}n, and G. J. de
Valc\'{a}rcel, Phys. Rev. Lett. \textbf{100}, 203601 (2008).

\bibitem {Lassen09}Related experiments have been started for the simplest case
($f=1$) we treated in \cite{Rotational}; see M. Lassen, G. Leuchs and U.L.
Andersen, arXiv: 0901.2783 (2009).

\bibitem {Drummond80}P. D. Drummond and C. W. Gardiner, J. Phys. A: Math. Gen.
\textbf{13}, 2353 (1980).

\bibitem {Carmichael}H. J. Carmichael, \textit{Statistical Methods in Quantum
Optics 1} (Springer, 1999).

\bibitem {Gardiner00}C. W. Gardiner and P. Zoller, \textit{Quantum Noise}
(Springer, 2000).

\bibitem {Collet}M. J. Collett and C.W. Gardiner, Phys. Rev. A \textbf{30},
1386 (1984).

\bibitem {Notation}Note that $\chi\alpha_{0}$ in \cite{Rotational} reads
$\gamma_{s}N$ in our notation.

\bibitem {nota}It is worth remarking one point on the diagonalization of
operator $\mathcal{L}_{l}$. Its eigensystem consists of two degenerate
subspaces. How must the two eigenvectors expanding each degenerate subspace be
chosen? As will become clearer later, the more appropriate choice is the one
that makes the operators related to the projections $c_{j}$ be either
Hermitian or anti-Hermitian, so that they coincide (up to a real or imaginary
constant) with some observables of interest. For example, with our choice of
eigenvectors we see that the operators related to $c_{1}^{l}$ and $c_{2}^{l}$
are $\hat{c}_{1,2}^{l}=\hat{a}_{+l}\pm\hat{a}_{+l}^{\dag}\pm\hat{a}_{-l}%
+\hat{a}_{-l}^{\dag}$ [see Eq. (\ref{eigen})], which are Hermitian and
anti-Hermitian, respectively.

\bibitem {Hodgson}N. Hodgson and H. Weber, \textit{Laser resonators and beam
propagation} (Springer, 2005).

\bibitem {Arnold}L. Arnold, \textit{Stochastic differential equations theory and applications} (Wiley, 1974).

\end{thebibliography}
\end{document}